\documentclass[pagesize=auto,bibliography=totoc,twoside=false]{scrbook}

\usepackage{comment}
\usepackage{etoolbox}
\usepackage{fancyvrb}
\usepackage{graphicx}
\usepackage[round]{natbib}
\usepackage{xspace}
\usepackage{xcolor}
\usepackage[hidelinks]{hyperref}
\usepackage{natbib}
\usepackage[font=small]{caption}

\usepackage[T2A]{fontenc}
\usepackage[utf8]{inputenc}
\usepackage[greek,russian,english]{babel}

\newcommand\nextpage[1][]{
\ifdefined\HCode {
  \HCode{<mbp:pagebreak />}}
\else
  \newpage
\fi
}

\def\surl#1{{\small{\url{#1}}}}

\ifdefined\HCode{\KOMAoptions{twoside=false}} \fi

\begin{document}



\nextpage

\chapter*{Exchange, obligation, accountability: \\ Moral orders of technology repair in Kampala, Uganda}

\newcommand{\chapterauthors}[3]{%
  \vspace{0.5em}
  {\large\bfseries #1} and {\large\bfseries #2}\\[0.3em]
  {\small #3}
  \vspace{1.5em}
}

\chapterauthors{Daniel Mwesigwa}{Steven J. Jackson}

As a long line of work in repair studies makes clear, repair is negotiated order \citep{graham_out_2007,jackson_rethinking_2014,henke_negotiating_2019}. It occurs through distributed action across networks of people and things always in flux and breakdown that are \textit{sometimes renormalized} and restored to working status through often invisible practices of maintenance and repair \citep{houston_values_2016}. Accordingly, to repair is not to fix \textit{something} but \textit{many} things at once: devices, systems, and infrastructures, but also people, relations, communities, even entire social orders \citep{henke_mechanics_1999,graham_out_2007,jackson_rethinking_2014}. Armed with this insight, studies of maintenance and repair have expanded dramatically in recent years and drawn new attention to the complex nature of action and meaning in repair worlds, from the importance of repair as a kind of counter-story to ‘innovation’ \citep{vinsel_innovation_2020}, to novel forms of skill and improvisation \citep{ahmed_learning_2015}, to the intersection of repair with wider practices and values of care, justice, and sustainability \citep{corwin_repair_2021,denis_soin_2022}. These studies have also produced new perspectives into work across sites from traditional repair desks to hobbyist hacker spaces, urban fabrication labs to rural open-air workshops, and contributed to more balanced and global stories of technology and infrastructure by calling out sites, skills, and locations long neglected within mainstream social and political conversations. In the process, they have generated new and critical insights around values in repair, and the complex sociocultural, political, and economic dynamics that connect and embed repair work in wider infrastructures, social systems, and worldviews. 

But while these studies have produced real and interesting insights into the nature, importance, and detailed practice of repair work, there has been less sustained attention to the greater and lesser norms and rules that shape and sustain action, practice, and meaning in repair worlds: the “structures of feeling” \citep{williams_marxism_1977,sharma_structures_2015} in which individual values and conduct are anchored, and the coherence of repair in collaborative life is sustained. If repair is indeed \textit{material} and \textit{practical} (as other contributors to this volume have shown), it is also \textit{normative} and \textit{ethical}, occurring within a landscape of interaction and exchange that is shaped and defined by values and expectations at least partly built into the ‘set-up’ of collaborative work. This includes shared (enough) judgments of good or bad action, generous or ungenerous, fair or unfair, upstanding or shady. If repair has and relies on a \textit{physical} set-up grounded in stuff and things, it also has an \textit{ethical} set-up, grounded in norms, rules, and values.

\section*{Moral orders of repair}

Moral orders of repair, as we define in this chapter, refer to \textit{specific} norms, rules, values, and expectations embedded in wider and historically contingent settings that structure and support joint work and exchange in repair worlds and other adjacent spheres of collaborative practice. This concept draws from two main lines of inquiry. First, a wide range of moral economy literature has illustrated that while customary rights and solidarity practices of exchange and redistribution have existed in pre-market societies \citep{thompson_moral_1971,polanyi_great_2010}, they have often come to the fore in moments of transition, as prior rights and relations are in process of being overwritten or undermined: for example, the transformation of land and labor into commodities introduced a new economic logic of a price-driven and wage economy, while also disembedding social life from market society. At the same time, as \cite{fassin_economies_2009} has argued, affect-laden “legitimizing notions” such as protests and ordinary forms of resistance were used by the aggrieved to confront \textit{modern} economic orders that had disrupted the lives of individuals and collectives. As such, moral economies are balanced systems in which the local structuring and justification of value (and valuation), along with the production and circulation of affective and emotional forces, guide but do not strictly determine moral judgment and action. Values and emotions in moral economies are thus not universal but deeply shaped by changing local and cultural contexts. While moral economies can motivate social and political action around concerns such as dignity and propriety, they can also be invoked to justify selfish and exploitative practices both in market and non-market contexts.

A second line of inquiry can be found in the growing philosophical and historical rethinking around the nature and function of ‘rules’ in social life. According to philosopher Charles Taylor \citeyearpar{taylor_philosophical_1997}, rules are not an abstraction of human action; while some rules are formulated, they are generally embodied (“values in flesh”): “serving as patterns of reasons for action.” This dual operation of rules – both as \textit{institutions} where rules are anchored and structured, and as \textit{embodiment}, through which rules are felt and experienced – guides “how an agent can understand a rule, be guided by it, without having an inkling of a whole host of issues that must be (it would appear) resolved” (p.~179). Beyond the abstract and enacted ideal of rules, Giorgio Agamben \citeyearpar{agamben_highest_2013} examines the expression of rules not only in governing behavior but also in constituting and reflecting an entire way of life. Based on medieval monastic rules, like those of St. Benedict and St. Francis, these rules provided examples through which life could be lived, while also establishing exceptions and provisos, in part premised on practice “absolutely outside of the determinations of the law” \citep[p. 110]{agamben_highest_2013}. This dialectic between rule-as-prescription and rule-as-form-of-life is what Agamben has suggested is key to monastic life, defined by a reciprocal relationship between rules and practice, where life itself transcends the law but remains shaped by it. 

Recent and established scholarship in Africanist economic anthropology and social theory extend and deepen these lines of thought. Jane Guyer’s \citeyearpar{guyer_marginal_2004} landmark work, \textit{Marginal Gains}, showed how monetary transactions in Atlantic Africa took different forms, including across multiple scales and valuation regimes inflected by centuries of slavery, trade, commerce, and encounters with “modernity.” Guyer has shown how “monetary transactions” are not limited to hard currencies and commodities, but may also account for transformations disciplined and performed through popular conventions and practices of exchange. The economic concept and popular practice of “marginal gains” can thus be understood as referencing and enacting profit-driven “purposive behavior”; but also anticipation; hope; and potential cracks or breakdowns in the interfaces of exchange and conversion. Through such processes, “individuals are thought to gain access to and control over people and resources through their mastery over transformative processes” (p.~39). Her famous telling of price setting during a fuel crisis in Nigeria in the 1990s foregrounds the performative work through which fuel (the lifeblood of the economy) was rationed and sold at unequal prices nevertheless widely accepted as “fair.” Following Guyer, Bize’s \citeyearpar{bize_right_2020} ethnography of roadside fuel trade practices in Kenya details local modes of “exceptional valuation” to tell us how marginalized communities engage with market relations and stake moral and practical claims for redistributive purposes. Drawing on the perspectives of truck drivers and fuel dealers, Bize shows how driver claims to the “right to the remainder [of leftover fuel]” serve to justify and moderate relations within the fuel trade while articulating (and subtly advancing) broader visions of economic justice and social order.

Beyond their immediate practice and efficacy, rules also endorse certain activities as appropriate and legitimate while driving others to the margins. Jennifer Hart’s \citeyearpar{hart_making_2024} book, \textit{Making an African City}, shows how centuries-old practices of commerce and market exchange in cities like Accra, Ghana were rendered “informal” by colonial structures and technocratic imaginations of the late 19th century. She details the steps by which ‘informality’ and ‘informal economy’ emerged as historically grounded terms rooted in “modernist rhetoric about the urban poor” (p.~22). Citing the promulgation of the Town Council Ordinance in colonial Ghana, Hart shows how the Ordinance’s obsession with universalist ideals of order was accompanied by the technocratic construction of ‘disorder’ and the problematization of African bodies and practices through epithets such as ‘nuisance,’ ‘illegal,’ and ‘pirate.’ Indeed, before these new kinds of regulatory actions, “market trading, corn milling, and other activities were not part of an `informal economy'; they were \textit{the} economy” (p.~23, emphasis in original). Yet in Accra (and almost certainly in cities across Africa), the locals adopted practices “outside of the state-sanctioned structures of social and economic life” as a kind of continued investment in self-determination in the “absence of accessible and meaningful alternatives” \cite[p. 23]{hart_making_2024}.

Taking these insights together, moral orders of repair have deep roots in the cultural and normative \textit{but are expressed primarily in practice}, through the everyday negotiations by which repair work is moderated, including through the management of ongoing breakdowns and achievement of value(s) in repair and collaboration \citep{houston_values_2016}. While moral orders can help us apprehend and read into \textit{macro} breakdowns in systems of exchange and the implicit rules that undergird them (for example, the moral and practical work of averting a fuel crisis in Guyer’s case), moral orders are also visible in everyday moments of \textit{micro}-crises in ordinary life and practice. Accordingly, the concept of rules extended in this case adopts a stance that blends rule-as-prescription and rule-as-form-of-life, linking formal definitions of rules to their informal and pragmatic expression, through the everyday practice of ‘expert’ or ‘novice’ participants. In this way, moral orders of repair provide a detailed specification and description of informal rules, norms, and values, which are gestured as implicit in moral economies. These specific norms, rules, and values inform complex and dynamic processes of repair while structuring actor expectations over time. 

Building on fieldwork conducted between 2022 and 2024 with repair workers in Kampala, Uganda, we argue that everyday repair practices are embedded in informal but organized systems of exchange, obligation, and accountability. Across two rounds of ethnographic research involving interviews, observations, and interactions in mobile phone and computing repair markets of downtown Kampala, we traced how actors articulated and navigated moral and normative claims around their work and life, while also drawing on unique forms of technical skill, experience, and expectations both within and beyond the repair shop.

\section*{Mapping repair worlds and repair actors}

The cast of repair worlds in downtown Kampala comprises four main kinds of actors. \underline{Repairers (or technicians)} work on a range of computing devices, notably mobile phones, laptops, and gaming consoles, and can be further sub-categorized as \textit{specialists} who work on specific high skill and high profile tasks such as “motherboard repair” or “advanced software repair”; \textit{generalists} who perform mid-range tasks such as fixing broken integrated circuit chips but mostly work on basic tasks like replacing broken screens and faulty batteries; and \textit{assemblers} who almost exclusively replace basic parts and accessorize devices. Large- and small-scale \underline{suppliers} ship products such as spare parts, accessories, and new or refurbished devices from China and Dubai to downtown Kampala. \underline{Middlemen}, popularly known as ‘abayiribi,’ a colloquial Luganda term indicating a certain quality of ‘street smarts,’ act as go-betweens, linking clients to repairers or suppliers. They earn primarily via commission and are often strategically positioned at busy nodes like building corridors and street corners, where they court clients and usher them to repairers or suppliers (and in isolated cases, scam them). Finally, \underline{clients} comprise the regular or occasional customers who visit repair or supplier shops to access a variety of repair goods and services. Some of these arrive as one-off clients steered by middlemen. However, others develop long-lasting relationships with repairers or suppliers whom they regularly visit for services and also recommend to others.

These categories are not entirely fixed, and boundaries between actors may be drawn and renegotiated in several ways. Actors like repairers and suppliers draw on their direct relationships to work in close proximity to each other, establishing informal but often enduring relations around things like the sourcing of necessary parts and referring clients, and occasionally the repurposing of serviceable parts from broken and unfixable phones. In some cases they switch roles, with repairers doubling as suppliers (especially of popular spare parts such as screens and batteries) and suppliers employing in-situ repairers to help with low-level repair work such as screen replacements. At certain moments, repairers and suppliers may also act as middlemen, as when they solicit repair products and services from elsewhere and keep the ‘margin’ in client fees. Clients too can receive commissions (usually disguised as tokens of appreciation) when buying products and services on behalf of colleagues or friends. Despite this role fluidity, however, some mobilities across roles can be challenging given the levels of skill required; protection or policing of self-interest, status, and reputation; and the expectations of past relationships. For example, it is rare for a \textit{bona fide} middleman to become a repairer without the necessary training and expertise, acquired in general through extended periods of informal apprenticeship (almost always with more established and better networked and reputed peers). While it is possible for, say, a supplier to consider a repairer as a ‘client,’ often a repairer and supplier’s pre-existing relationship shapes sets of shared obligations and expectations. Within this partially fixed, partially fluid world, value (and sometimes money) is constantly changing hands, according to rules and conventions that are structured but rarely fixed or fully specified in advance. Action and practice in these worlds is illustrated in three key dimensions that particularly stood out in our work – fair exchange, collaboration across hierarchy, and relational accountability – which taken together organize the moral order and collective practice of repair in the mobile phone and computing markets of Kampala.

\section*{Fair exchange}

A significant part and aspect of transactions in repair worlds consists of fixing “good prices.” But what constitutes a good price, considered fair by all actors in a transaction? Several interlocutors suggested that whereas clients drew on repertoires of knowledge and relations to get what they considered a good price, more often than not pricing repair products and services was in the hands of the actor coordinating a deal; a repairer, supplier, or middleman. Judgments of fair price shape the interactions of repairers and the various intermediaries: if a repairer has an established relationship with a middleman, they might charge lower with the expectation that the middleman will also take a commission. Below, we detail an exchange(s) capturing how a ‘fair’ price for a repair service is made, following the unfolding of negotiations, decisions, and backgrounded actions that played a role in the determination of a final price.

\begin{quote}
    \textit{A client, a middle-aged man, broke his Motorola E20 smartphone screen earlier in the day. Needing a screen replacement, he visits Xavier, a repairer whom he has known for years. Xavier, an assembler who typically handles simple repairs, accepts the job and casually passes the phone to his assistant, Xander, saying only one word: “screen.” No further explanation is needed. Without delay, Xander and I set off into the dense commercial labyrinth of downtown Kampala, where repairers and suppliers operate in tightly packed alleys and multistory buildings.}
    
    \textit{    We visit several suppliers, most of whom do not stock Motorola screens. At one shop, a supplier examines the phone and murmurs a quote of Ugx 150,000 (around US\$40). Sensing discretion is needed, Xander quickly clarifies that I am not the client. He rejects the offer, calling it too high, and updates Xavier by phone as we return. At the repair shop, Xavier is mid-conversation with the client, quoting a price of Ugx 200,000 (\$54) for the job. The client hesitates, prompting a performance of price negotiation. Xavier calls Manda, a known supplier, and soon offers a revised price of Ugx 170,000 – broken down into Ugx 150,000 for the spare part and Ugx 20,000 for labor. It appears like a gesture of compromise. The client agrees.
    } 

      \begin{figure}[htbp]
  \centering
  \includegraphics[width=0.6\textwidth]{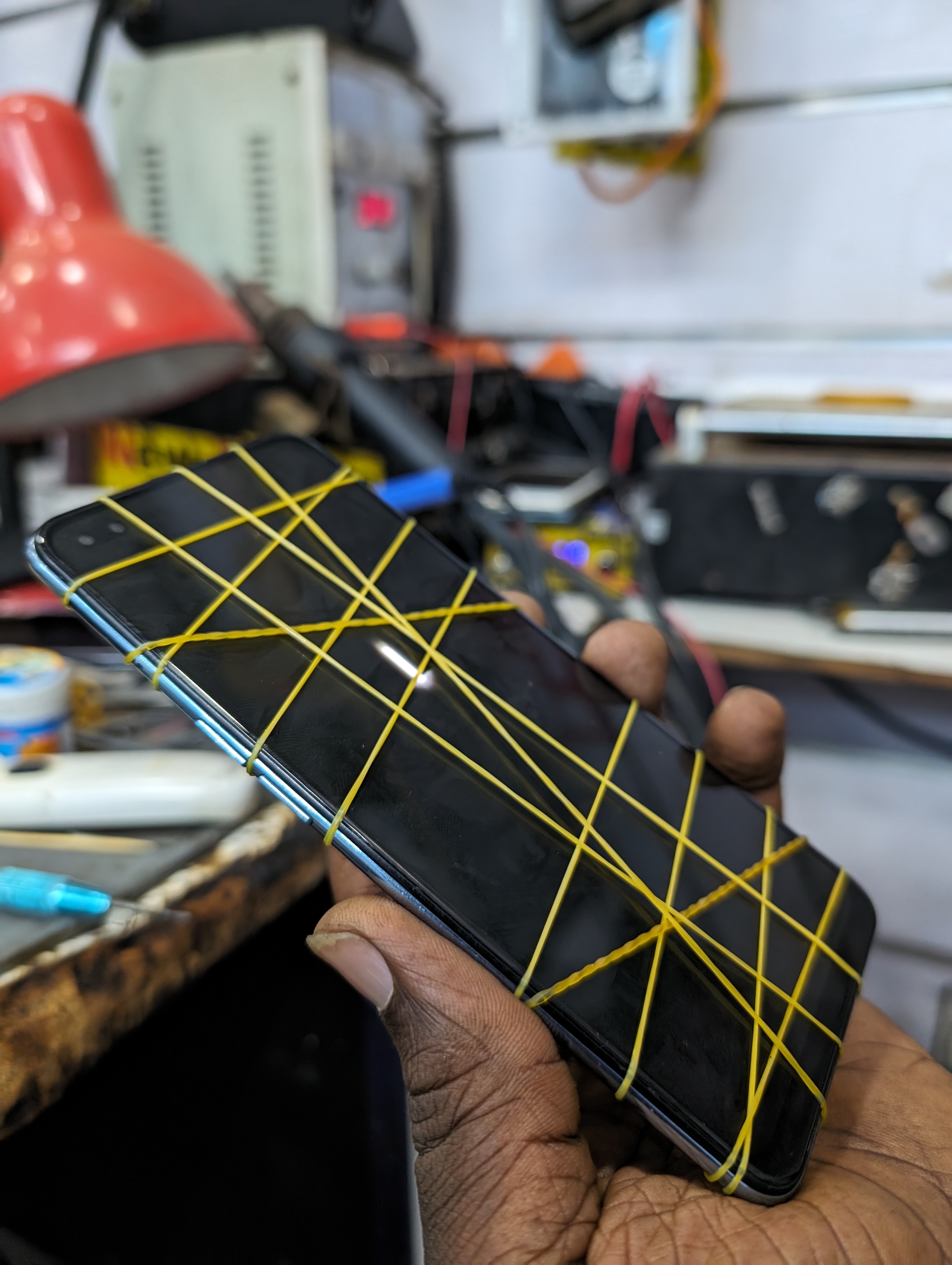}
  \caption{A newly fixed screen, securely tied with rubber bands to enable the screen replacement to stick on the phone’s frame faster.}
  \label{fig:yourlabel}
\end{figure}
    
    \textit{Xander heads off to Manda’s, where the screen is sourced and installed. The phone is returned, seemingly good as new. The client appears satisfied with the service and the final price. But much has unfolded backstage. Xavier had secured the screen for just Ugx 70,000 (\$19). The margin, more than double, is built into the final cost. Yet the transaction feels fair to the client. As Xander later sums it up, “The client feels like he was not cheated.” 
    }

\end{quote}

There are two key insights from this encounter that speak to how fairness in pricing is negotiated and accomplished. First, repairers like Xavier and Xander control the pricing process through a mix of social and technical skills arriving at a final price perceived as fair but still economically advantageous to them. While the repairers worked in tandem to establish a price for their client, the ultimate goal was to ensure that the client did not feel like he was cheated. In managing the client’s expectations and feelings of fairness, Xander and Xavier leveraged their networks, pricing skills, and intuitions, honed through years of experience. Ahmed et al. \citeyearpar{ahmed_learning_2015} emphasize that this sort of social knowledge of repair comprises a key aspect of “how [repairers] deal with customers and fix repair service fees.” Second, clients actively draw on bargaining skills and personal networks to advocate for themselves, producing a sense of fairness and satisfaction. Fair exchange, where it occurs, emerges from a balance between client needs and expectations and repairers’ efforts to manage those expectations amid limited client knowledge. Clients often lack both the technical skills and the insider understanding of repair worlds, while repairers rely on social and technical expertise, along with collaborators, to shape both the material and discursive aspects of repair. In this transaction, back channels play a central role: a) Xander’s debrief call to Xavier about an earlier supplier quote, which the client is not privy to, and b) Xavier’s call to Manda, where he learns the screen actually costs Ugx 70,000 but presents the price as reduced to Ugx 150,000 for the client’s benefit. These moments show how repairers expertly manage the flow of information to maintain credibility, allowing them to profit while still delivering a price the client is willing (and happy) to pay. Beyond client-repairer relations, expectations of fair exchange were also common amongst repairers and between repairers and middlemen, who recognized shared aspirations and mutual obligation. Ultimately, fair exchange aspires to a more horizontal relation where real or perceived differences between repair actors are flattened to generate a sense that exchanges and transactions are conducted on even grounds.

\section*{Collaboration across hierarchy}

Besides monetized performances that characterize repair worlds, practice in these worlds is also subtended by non-monetary actions and obligations: performances and expectations that nourish “the brotherhood,” as put by Pius, a specialist smartphone and laptop software repairer. These practices are elicited at moments of need but are also frequently mundane, essential to the sustenance of networks and relationships in repair worlds. While doing good includes looking out for each other, sharing necessary and appropriate resources – tools, food, time, specialty skills, etc. – there are also tensions in practices of sharing and collaborating across hierarchy. Below, we share a series of encounters at Daliwo and Muki’s repair shop. Muki is a specialist iPhone motherboard repairer who mentored Daliwo into the same trade: Daliwo has since scaled the ranks as one of the most distinguished smartphone motherboard repairers in town.

\begin{quote}
    {\em Daliwo and Muki’s repair shop bustles with activity. Amidst interruptions and incessant phone calls from middlemen (who dominate their clientele), the repairers tend to their work with dexterity and care. Seated before RL-M3T microscopes, their eyes stay fixed to the lenses as their hands work on iPhone motherboards beneath, occasionally referencing schematics on old laptops by their sides. Their routines seem almost choreographed, alternating between soldering irons in one hand and tweezers in the other. The iron heats specific points on a motherboard while the tweezers hold tiny components as they are moved. Daliwo works on an iPhone 13 motherboard, trying to fix its failure to relay voice (“amaloboozi” in Luganda). He carefully solders a wick, a thin metallic conductor, around the motherboard's edges to sharpen them for stronger contact with other components during reassembly. Smoke billows under the microscope. The iron and wick are cooking. But everything seems under control.
    }

    \begin{figure}[htbp]
  \centering
  \includegraphics[width=0.8\textwidth]{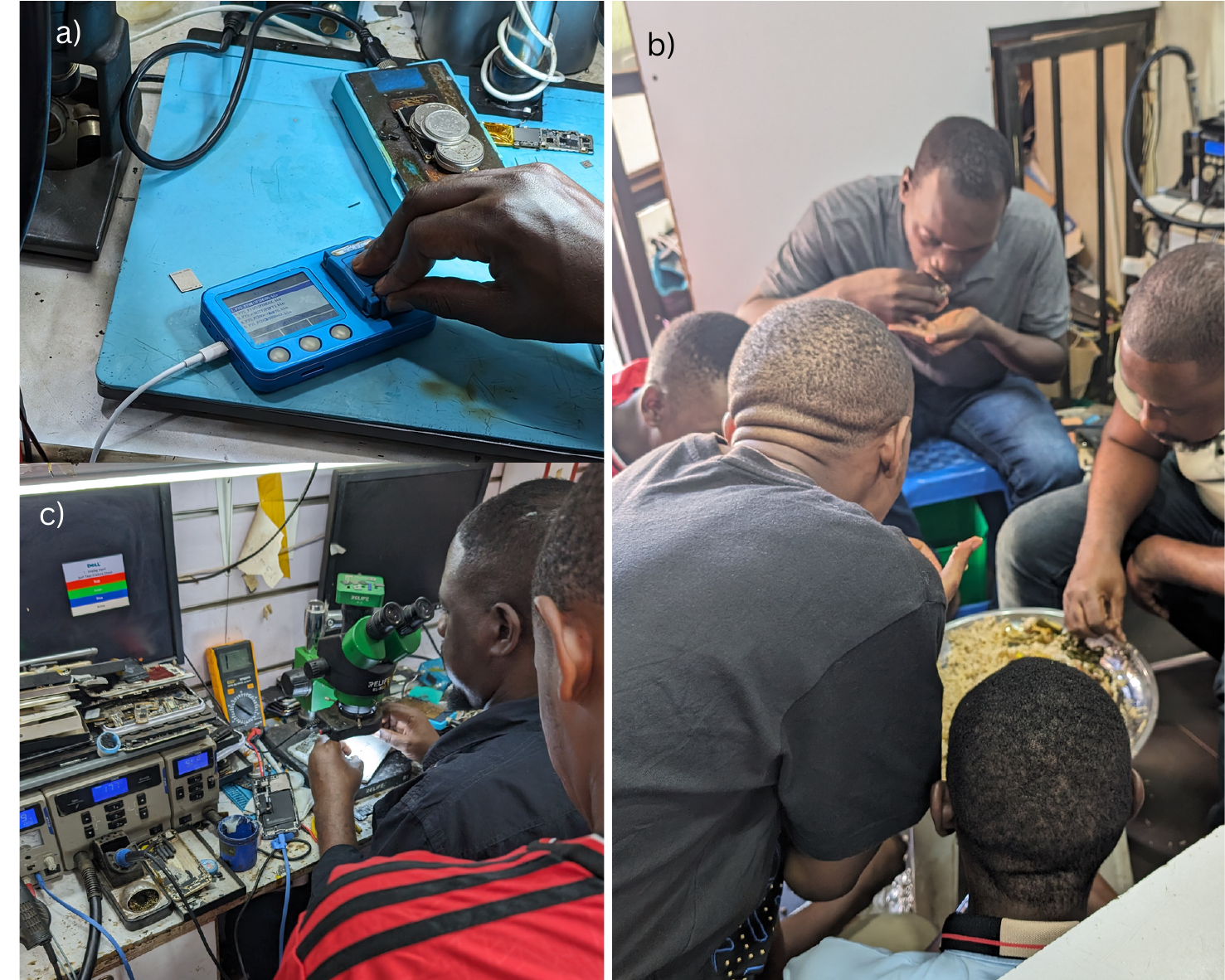}
  \caption{(a) A blue NAND programmer. (b) Repair actors sharing \textit{lusaaniya} (a dish of food) in a crammed repair shop. (c) A specialist repairer working a microscope, overlooked by an apprentice.}
  \label{fig:sharing}
\end{figure}

    {\em In their cramped repair room, Daliwo and Muki engage with a steady stream of people and activities, often middlemen observing repairs or clients calling to follow up. A client brings in a Google Pixel 2 that will not power on. Though Daliwo specializes in iPhones, he opens the Pixel and tests it with a multimeter: no positive result. He does this free of charge. Someone asks for help using a P7S NAND programmer, a pocket-sized tool used to transfer data between memory chips. Daliwo, familiar with the person’s boss, offers quick instructions, describing it as plug and play. When the chip resists reprogramming, he advises cleaning it and offers the tool for free, suggesting it will work later. As calls continue to come in – some seeking advice, others checking on repair status – Daliwo also makes follow-up calls, like checking if the NAND chip data transfer succeeded. Amidst it all, work on the iPhone 13 motherboard resumes.
    }

\end{quote}

The above story highlights the importance of collaboration across hierarchy, where repairers draw on and extend non-monetary practices to sustain social, cultural, and economic ties across the wider repair community, albeit not without tension. We see first how repair actors leverage their resource pools to nourish and facilitate the work of collaborators and colleagues (even, sometimes, competitors). By using expensive equipment such as microscopes and NAND programmers, Muki and Daliwo shared specialized skills and time and attention with clients, predominantly middlemen and other repairers, performing quick diagnoses for their devices while also advising on technical fixes and pricing mechanisms. They did this frequently free of charge, while also appearing eager to help or mentor those willing to learn. Through this work and through their distinct forms of skill and access to key equipment, they also assumed important and widely acknowledged leadership roles in the community. Second, while Muki and Daliwo went out of their way to provide help to their clients, some repairers and middlemen expressed an entitlement to free services and real-time access that was not reciprocated,  resulting in repeated and late night calls and delays in promised or expected work that provoked hard feelings, verbal altercations, and subtle reworkings in the calculus of deserving and undeserving clients.

Taken together, these acts of collaboration are personalized by the leading actors, structuring hierarchies as horizontal relations meant to flatten differences between the leading actors and their collaborators (although it is apparent in repair worlds studied that hierarchies are vertically ordered). Beyond any individual relations, our actors (including Muki and Daliwo) echoed the need to sustain “the brotherhood,” the loose collective of various repair actors on which they all ultimately relied and who in the future might be needed for help or a favor. These practices also involved navigating and resolving tensions and breakdowns in both the social and technical dimensions of repair work and were widely understood as endemic to a shifting and uncertain field. Sometimes this came at non-trivial personal and social costs, including working long hours and bearing additional personal economic costs to sustain the community of (other) repair actors. In our fieldwork, these obligations were borne regularly and formed an important baseline of function, exchange, and consistency that kept the wider repair enterprise going.


\section*{Relational accountability}

The art of navigating repair worlds depends on how actors manage the contexts in which their work is embedded, including how they deal with sometimes surprising and challenging outcomes. As Muki, a specialist repairer conveyed, it is not uncommon for electronic gadgets to “die as a patient would in a [hospital] theater.” Such accidents, which are part of the work that constitutes repair, are most commonly resolved directly with clients (during which moments, reservoirs of trust built through reputation or prior relation may prove critical). In other cases however local authorities and the police may be enlisted to adjudicate such matters. In the story below, we show an unexpected fieldwork encounter where Jama, a middleman working for a repairer and an optician, was arrested for performing ‘illegal’ repair work on the streets of Kampala, only achieving his freedom by appealing to one of his \textit{employers'} sense of responsibility.

 \begin{figure}[htbp]
  \centering
  \includegraphics[width=0.8\textwidth, trim=0 20 0 20, clip]{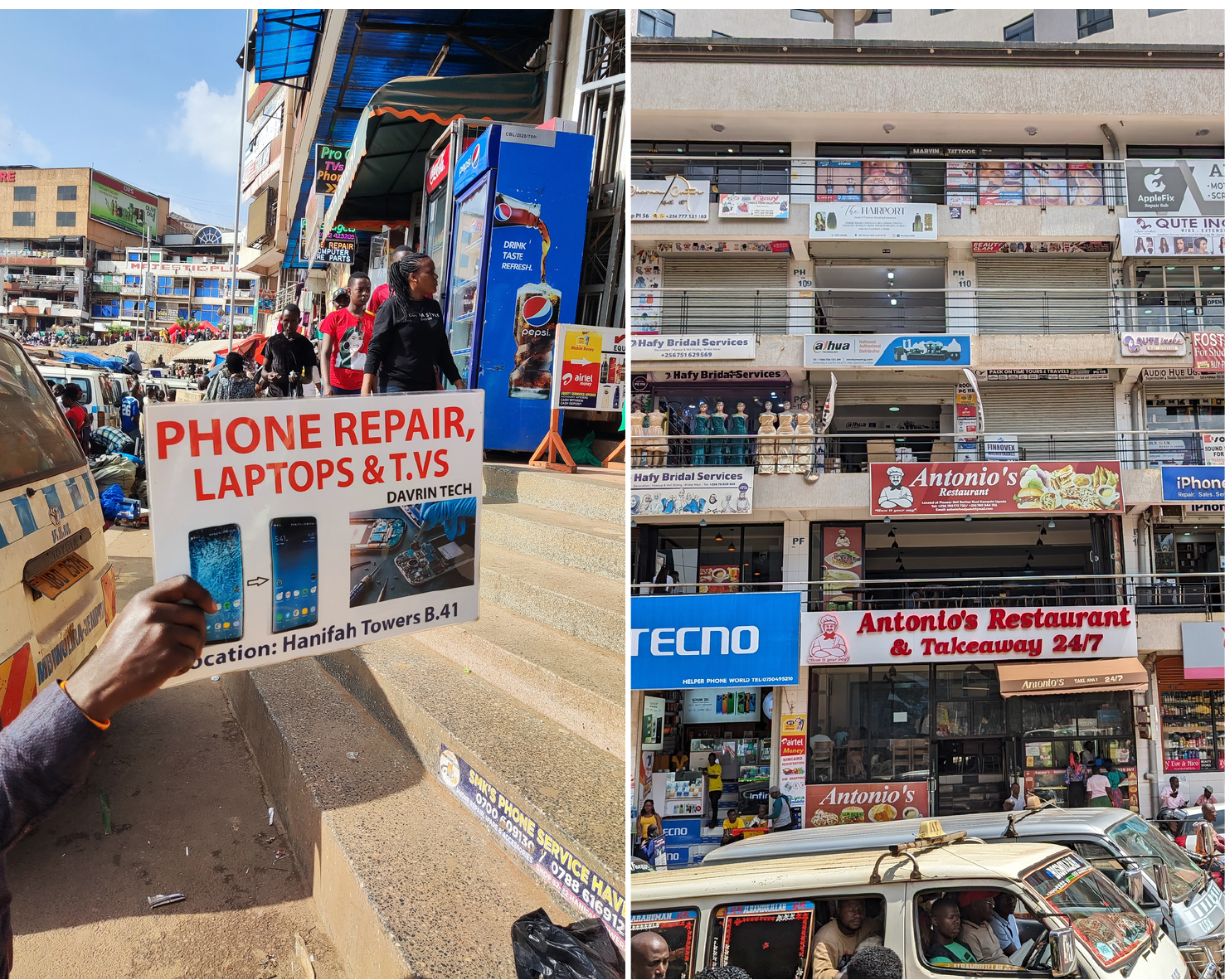}
  \caption{On the left is a hand of a man holding a ‘phone repair’ signpost. On the right is a multi-storied building in downtown Kampala, adorned with multiple signposts.}
  \label{fig:signpost}
\end{figure}

\begin{quote}
    {\em It’s a hot afternoon in downtown Kampala, and Jama stands on the street holding a large embossed signpost that reads “phone repair.” Though the building behind him is covered in signage of all kinds, it is Jama’s presence that draws people in. Trained in phone repair through a local youth program, Jama prefers this commission-based signpost work, which he does for up to ten hours a day to earn “akawogo” (colloquial term for money in Luganda). He also carries a second signpost advertising eye care services in the same building, earning a little extra. Such sign work is illegal (though widespread), and run-ins with city council enforcement agents are common. Just after recounting how a friend had been arrested earlier that day, Jama is suddenly grabbed from behind by an agent and arrested. When questioned, the officers offer no clear reason, first saying he has done nothing wrong, then charging him with “idleness,” a colonial-era offense historically used to police movement and control informal urban life.
    }
    
    {\em Jama is dragged through the city to a nearby police post, an aluminum kiosk with a dimly lit holding cell. On the way, the agents confront other street vendors but arrest no one else. Jama is their main focus, and while he begs the agent for forgiveness, this comes at a cost. After twenty minutes in the cell, a plainclothed officer escorts Jama to a mobile money kiosk, where he withdraws money to pay for his release: a bribe of Ugx 45,000 (\$12), an amount more than twice that he makes per day. This payment is not an ‘official fine’ but greases the hands that expiate him from being committed to the courts of law; it facilitates his freedom. Jama explains that his boss at the repair shop had stopped taking his calls, claiming there were no earnings that day to cover bail. Left with no other choice, Jama called the optician he also advertises for, who unexpectedly sent him the money, even though they had no such prior agreement and little prior relation. By the end of it all, Jama summed up the day with a shrug: “This is Uganda.” It was a gesture of resignation, but also one of persistence. Whether he is culpable or not, life must go on, and he will try again the next day.}

\end{quote}

Two key insights emerge from this encounter. First, the practice of repair is uncertain, and its consequences are personal. Jama’s arrest was a reminder of the precariousness of repair work, perhaps especially in its most exposed variant in the streets. However, it also showed how he overcame his predicament (including the no-show by his repairer boss) by enlisting the help of his secondary boss (the optician), whose sense of responsibility he appealed to. Second, the case shows the complex interplay between repair worlds and more formally constituted authorities, including in their most immediate and often challenging form, the police. Police were drawn into other instances in our study, including mediating disputes or overseeing compensation over failed repairs, in most cases exacting a similar kind of ‘fee’ for their services. These dynamics reflect a form of accountability that is neither institutional nor guaranteed, but contingent, relational, and ultimately dependent on the recognition (or not) of relations by more powerful actors, whether patrons or police. Jama’s case shows both the precarity of informal practice and the material and discursive strategies by which individuals in weaker positions mobilize against those above them -- with whom they share social and collaborative relations -- to achieve security and redress.


\section*{Conclusion}
In this chapter, we have developed the concept of moral orders of repair, defining it as the norms, rules, values, and expectations that support and structure practice, exchange, and collaboration in everyday repair worlds and beyond. To demonstrate the utility of the concept, we charted the social, cultural, and economic conditions in which moral orders emerge, connecting diffuse activities, sites, practices, and actors, and how growing tensions in collaborative life and practice are distributed and managed. Under the ever-present conditions of breakdown, moral orders may thus provide resources to explain and justify action (or non-action): for example, the dimensions called out in the practices of negotiating prices considered ‘fair’ by all actors; the decision to take on fixes for the brotherhood ‘for free’; and whether or not to bail out a loosely connected signpost holder. Accordingly, moral orders are not a standalone property or state of being – of things or of people – but provide structure or scaffolding for both practice and relations. While moral orders are expressed in moments of exchange, they exist and persist well beyond them, including longer anchorings in place and history (like the colonial echoes of a ‘law-and-order’ arrest on a Kampala street corner). Moral orders are therefore where the rubber meets the road, and moral economies are made real and enter into the flow of collaborative practice and exchange. 

By lingering on the nature of action within and beyond repair worlds studied, we foreground the hidden rules and practices that both constitute and reshape the worlds from which moral orders emerge. Because moral orders are concrete and specific (evinced in the three key dimensions), they go beyond the ‘implicit rules’ invoked in mainstream humanities and social sciences scholarship, potentially guiding how maintenance and repair studies can account more concretely for the rules that undergird action amongst individuals and collectives, and the beliefs and structures that mold that action. Moral orders also reflexively reproduce and shape spheres of practice and life. Beyond repair, moral orders speak to the \textit{how} of ethics and practice: how opposing concerns in work are navigated and balanced. They also speak to the \textit{why}, showing why things are the way they are while providing the connections to understand action across seemingly disparate sites. In this way moral orders situate action, binding it in time and place. They are where action unfolds into the world, and the world pushes back. 




\bibliographystyle{apalike}
\bibliography{references}

\end{document}